\documentclass[12pt]{article}
\usepackage{graphicx}
\usepackage{dcolumn}
\usepackage{bm}
\usepackage{color}
\usepackage{amssymb}
\usepackage{amsmath}
\usepackage{amsthm}
\usepackage{natbib}
\usepackage{multicol,times}
\usepackage{multirow}
\newcommand{\df}{\mathrm{d}}
\newcommand{\tc}{\tilde{t}_c}

\newtheorem{proposition}{Proposition}
\newcommand{\E}{\mathbb{E}}

\begin{document}

\title{\bf Diagnostics of Rational Expectation Financial Bubbles with Stochastic Mean-Reverting Termination Times} 
\author{L. Lin{\small$^{\mbox{\,\ref{ETH}, \ref{BUAA}~\footnote{Email: llin@ethz.ch}}}$} and D. Sornette{\small$^{\mbox{\,\ref{ETH}, \ref{SFI} \footnote{Email: dsornette@ethz.ch}}}$} 
}%
\date{}

\maketitle

\vspace{-9mm}
\begin{enumerate}
\item{Chair of Entrepreneurial Risks, Department of Management, Technology and Economics, ETH Zurich, Kreuplatz 5, CH-8032 Zurich, Switzerland}\label{ETH} \vspace{-3mm}
\item{School of Economics and Management, Beihang University, 100191 Beijing, China \vspace{-3mm}%
\item{Swiss Finance Institute, c/o University of Geneva, 40 blvd. Du Pont d'Arve, CH 1211 Geneva 4, Switzerland}\label{SFI}
}
\label{BUAA}%
\end{enumerate}

\vspace{1mm}

\vspace{1mm}
\begin{center}
\textbf{Abstract}
\end{center}
\begin{quote}
\hspace{0.5cm}
We propose two rational expectation models of transient financial bubbles
with heterogeneous arbitrageurs and positive feedbacks leading to 
self-reinforcing transient stochastic faster-than-exponential price dynamics. 
As a result of the nonlinear feedbacks, the termination of a bubble is found to 
be characterized by a finite-time singularity in the bubble price formation process
ending at some potential critical time $\tilde{t}_c$, which follows a
mean-reversing stationary dynamics. Because
of the heterogeneity of the rational agents' expectations, there is a synchronization problem
for the optimal exit times determined by these arbitrageurs, which 
leads to the survival of the bubble almost all the way to its theoretical end time. The explicit exact
analytical solutions of the two models provide
nonlinear transformations which allow us to develop novel tests for the presence
of bubbles in financial time series.  Avoiding 
the difficult problem of parameter estimation of the stochastic differential equation describing the price dynamics,
the derived operational procedures allow us
to diagnose bubbles that are in the making and to forecast their termination time.
The tests performed on three financial markets, the US S\&P500 index from 1 February 1980 to 31 October 2008, 
the US NASDAQ composite index from 1 January 1980 to 31 July 2008 and the
Hong Kong Hang Seng index from 1 December 1986 to 30 November 2008, 
suggest the feasibility of advance bubble warning.

\end{quote}
\hspace{0.5cm}
\begin{quote}
\end{quote}

\section{Introduction}

Bubbles and crashes in financial markets are of global significance because of
their effects on the lives and livelihoods of a majority of the world's
population.  While pundits and experts alike line up after the fact to claim
that a particular bubble was obvious in hindsight, the 
real time development of the bubble is often characterized by either a deafening silence
or a cacophony of contradictory opinions. Here, we propose
two models of financial bubbles, from which we develop the corresponding operational procedures
to diagnose bubbles that are in the making and to forecast their termination time.
The tests performed on three financial markets, the US S\&P500 index from 1 February 1980 to 31 October 2008, 
the US NASDAQ composite index from 1 January 1980 to 31 July 2008 and the
Hong Kong Hang Seng index from 1 December 1986 to 30 November 2008, 
demonstrate the feasibility of advance bubble warning on the major market
regimes that were followed by crashes or extended market downturns.
The empirical results
support the hypothesis that financial bubbles result from positive feedbacks
operating on the price and/or on its momentum, leading to faster-than-exponential
transients.

These results should be appreciated from the perspective of the present
state-of-art on modeling and detecting bubbles.
There is no really satisfactory theory of bubbles, which both encompasses its
different possible mechanisms and adheres to reasonable economic principles (no
or limited arbitrage, equilibrium or quasi-equilibrium with only transient
deviations, bounded rationality). Part of the
reason that the literature is still uncertain on even how to define a bubble is
that an exponentially growing price can always be argued to result from some
fundamental economic factor \citep{Gurkaynak2008,Lux2002}.  This
is related to the problem that the fundamental price is not directly
observable, giving no strong anchor to understand observed prices.
Another fundamental difficulty is to go beyond equilibrium to out-of-equilibrium
set-ups \citep{brock1993,brock1999,Chiarella2008,Hommes2008}.

Two conditions are in general invoked as being necessary for prices to deviate
from fundamental value. First, there must be some degree of irrationality in
the market. That is, investors' demand for stocks must be driven by something
other than fundamentals, like overconfidence in the future. Second, even if a
market has such investors, the general argument is that rational investors will
drive prices back to fundamental value. For this \textit{not} to happen, there
needs to be some limits on arbitrage. \citet{Shleifer1997}
provide a description for various limits of arbitrage. With respect to the
equity market, clearly the most important impediment to arbitrage are short
sales restrictions. Roughly 70\% of mutual funds explicitly state (in SEC Form
N-SAR) that they are not permitted to sell short \citep{almazan2004}. Seventy-nine
percent of equity mutual funds make no use of derivatives whatsoever (either
futures or options), suggesting further that funds do not take synthetically
short positions \citep{Koski1999}. These figures indicate that the vast majority of
funds never take short positions. Then, the argument goes that
bubbles can develop because prices reflect mainly the remaining optimistic opinions
and not the negative views of pessimistic traders who are already out of the market,
and who would take short positions, if given the opportunity.

One important class of theories shows that there can be large movements in
asset prices due to the combined effects of heterogeneous beliefs and
short-sales constraints. The basic idea finds its root in the original CAPM
theories, in particular, the model of \citet{Lintner1969} of asset prices
with investors having heterogeneous beliefs. Lintner and many others after him,
show that widely inflated prices can occur
\citep{Miller1977,Jarrow1980,Harrison1978,chen2002,Scheinkman2003,Duffie2002}.
In these models that assume heterogeneous beliefs and short sales restrictions,
the asset prices are determined at equilibrium to the extent that they reflect
the heterogeneous beliefs about payoffs.  But short sales restrictions force
the pessimistic investors out of the market, leaving only optimistic investors
and thus inflated asset price levels. However, when short sales restrictions no
longer bind investors, then prices fall back down. This provides a possible
account of the bursting of the Internet bubble that developed in
1998-2000. Many of these models take into account explicitly the relationship
between the number of publicly tradable shares of an asset and the propensity
for speculative bubbles to form. So far, the theoretical models based on agents
with heterogeneous beliefs facing short sales restrictions are considered among
the most convincing models to explain the burst of the Internet bubbles.

The role of ``noise traders'' in fostering positive feedback trading 
has been emphasized by a number of models.
For instance, \citet{DeLong1990} introduced a model of market bubbles and crashes which
exploits this idea of the role of noise traders in the development of bubbles,
as a possible mechanism for why asset prices may deviate from the fundamentals
over rather long time periods. Their work was followed by a number of
behavioral models based on the idea that trend chasing by one class of agents
produces momentum in stock prices \citep{barberis1998, Daniel1998,  Hong2005}.  
An influential empirical evidence on momentum strategies came from the
work of  \citet{jegadeesh1993, jegadeesh2001}, which established
that stock returns exhibit momentum behavior at intermediate
horizons. Strategies which buy stocks that have performed well in the past and
sell stocks that have performed poorly in the past generate significant
positive returns over 3- to 12- month holding periods. \citet{Debondt1985}
documented long-term reversals in stock returns. Stocks
that perform poorly in the past perform better over the next 3 to 5 years than
stocks that perform well in the past. These findings present a serious
challenge to the view that markets are semi-strong-form efficient.

It is important to understand what mechanisms prevent arbitrageurs from 
removing a bubble as soon as they see one. \citet{abreu2003} have 
proposed that bubbles continue to grow due to a 
failure of synchronization of rational traders, so that the later
choose to ride rather than arbitrage bubbles. \citet{abreu2003}  consider a market
where arbitrageurs face synchronization risk and, as a consequence, delay usage
of arbitrage opportunities. Rational arbitrageurs are supposed to know that the
market will eventually collapse. They know that the bubble will burst as soon
as a sufficient number of (rational) traders will sell out. However, the
dispersion of rational arbitrageurs' opinions on market timing and the
consequent uncertainty on the synchronization of their sell-off are delaying
this collapse, allowing the bubble to grow. In this framework, bubbles persist
in the short and intermediate term because short sellers face synchronization
risk, that is, uncertainty regarding the timing of the correction. As a result,
arbitrageurs who conclude that the arbitrageurs are yet unlikely to trade
against the bubble find it optimal to ride the still-growing bubble for a
while.

\cite{bhattacharya2008} provide a summary of recent efforts to
expand on the above concepts, in particular to address the two main questions
of (i) the cause(s) of bubbles and crashes and (ii) the possibility to diagnose
them ex-ante. Many financial economists recognize that positive feedbacks and
in particular herding is a key factor for the growth of bubbles. Herding can
result from a variety of mechanisms, such as anticipation by rational investors
of noise traders' strategies \citep{DeLong1990}, agency costs and monetary
incentives given to competing fund managers \citep{Dass2008} sometimes leading to
the extreme Ponzi schemes, rational imitation in the presence of uncertainty
\citep{Didier2000} and social imitation.  The bubble models
developed here build strongly on this accepted notion of herding.
We refer to \citet{Kaizoji} for an extensive
review complementing this brief survey.

The present paper takes its roots in the Johansen-Ledoit-Sornette (JLS) model \citep{JLS1,JLS2} 
formulated in the Blanchard-Watson framework of rational expectation bubbles \citep{blanchard1979,blanchard1982}.
The JLS model combined a representation of the herding behavior of noise traders
controlling a crash hazard rate  with the arbitrage response of rational traders
on the asset price. One implication of the JLS model is the transient faster-than-exponential
acceleration of the price due to the positive feedback associated with the herding
behavior of noise traders. This faster-than-exponential pattern can theoretically 
culminate in a finite-time singularity, which characterizes the end of the bubble
and the time at which the crash is the most probable. Other models have explored
further  the hypothesis that bubbles can
be the result of positive feedbacks and that the dynamical signature of bubbles
derives from the interplay between fundamental value investment and more
technical analysis. The former can be embodied in nonlinear extensions of the
standard financial Black-Scholes model of log-price variations
\citep{Didier2002,IdeSornette1,Corcos2002,Andersen2004}.
The later requires more significant extensions to account for the 
competition between inertia between analysis and decisions, positive momentum feedbacks
and negative value investment feedbacks \citep{IdeSornette2}.
Close to our present formulation, \citet{Didier2002, Andersen2004}
develop a nonlinear generalization of the Black-Scholes process which can be solved analytically.
The nonlinear feedback is acting as the effect of price on future growth, according to the
view that high prices lead to a wealth effect that drives behavioral investors to invest more aggressively.

The present paper adds to the literature by developing two related models of transient bubbles
in which their terminations occur at some potential critical time $\tilde{t}_c$,
which follows a mean-reversing stationary process with a fixed unconditional mean $T_c$. 
These models provides straightforward ways to determine the potential critical time without confronting 
the difficult problem of parameter estimation of the stochastic differential equation describing the price dynamics of price.
In our models, rational arbitrageurs can diagnose bubbles but do not know precisely when they end. 
These investors are assumed to form rational expectations of the potential critical time but not necessarily
of the detailed price process itself, which form a much weaker condition for investors' rationality.
Furthermore, we assume that rational arbitrageurs hold consistent expectation for the potential critical time with unbiased errors.  Although rational arbitrageurs know that the bubble will burst at its critical time, they can not make 
a deterministic prediction of this time and therefore of when other arbitrageurs will sell out, because they have little knowledge about others' belief about  the process governing the stochastic critical time $\tilde{t}_c$. Our rational investors
continuously update their beliefs on the probable termination of the bubble, according to their observation
of the development of the bubble. They exit the market by maximizing their expected payoff, based on their
subjective perception of the market bubble risk and the knowledge of the bubble dynamics. Because
of the heterogeneity of these rational agents' expectations, there is a synchronization problem
between these arbitrageurs, which 
leads to the survival of the bubble almost all the way to its theoretical end time. In this respect, our model 
is reminiscent to that of  \citet{abreu2003}, since the resilience of the bubble
results from the lack of synchronization between arbitrageurs on the decision to exit the bubble due to the heterogeneity
of their optimal exit times. Our models have the advantage of being quantitatively testable and their
concrete implementation provides diagnostics of bubbles in real time series, as we demonstrate below.

The first model, which leads to a finite-time singularity in 
the price dynamics with stochastic critical time is presented in the next section \ref{ththwrbw}.
This model generalizes Sornette and Andersen's model to allow for a mean reversal 
dynamics of the bubble end. Section \ref{th3wacvqtegbtt4} present a second model
leading to a finite-time singularity in the momentum price dynamics with stochastic critical time.
Both models exemplify the importance of positive feedback, which is quantified
by a unique exponent $m$. A value of $m$ larger than $1$ (respectively $2$) 
characterize a bubble regime in the first (respectively second) model.
The two models can be solved exactly in explicit analytical forms. These solutions provide
nonlinear transformations which allow us to develop novel tests for the presence
of bubbles in financial time series.  These two classes of tests, one for each bubble model, are 
developed respectively in subsections \ref{S: 4.1} and \ref{S: 4.2} and applied on
 three financial markets, the US S\&P500 index from 1 February 1980 to 31 October 2008, 
the US NASDAQ composite index from 1 January 1980 to 31 July 2008 and the
Hong Kong Hang Seng index from 1 December 1986 to 30 November 2008.
Section 5 concludes.

\section{First bubble model: finite-time singularity in 
the price dynamics with stochastic critical time \label{ththwrbw}}

Financial bubbles are often viewed as being characterized by 
anomalously high growth rates resulting from temporary
over-optimistic beliefs in a `new economy' or in a `paradigm shift' of the
fundamental structure of productivity gains. However, this definition
is unsatisfactory because a high growth rate associated with an 
exponentially growing price can always be justified by some other fundamental
valuation models which use higher discount factors and larger
dividend growth expectations, or introduce new accounting rules
incorporating for instance the benefits of real options. This definition
also leaves a large ambiguity as to when the bubble is supposed
to end, or when a crash might occur.

In contrast, we define a bubble
as a transient faster-than-exponential growth of the price, which would end in a 
finite-time-singularity in the absence of a crash or change of regime.
Such `super-exponential regime' results from the existence
of positive nonlinear feedback mechanisms amplifying past price increases
into even faster growth rates. These positive nonlinear feedback mechanisms
may be due to a variety of causes, including derivative hedging strategies, 
portfolio insurance methods
or to imitative behaviors of bounded rational arbitrageurs and of noise traders.
While herding has been largely documented to be a trait
of noise traders, it is actually rational 
for bounded rational agents to also enter into social imitation, as the
collective behavior 
may reveal information otherwise hidden to the agents. As a
result of the nonlinear positive feedbacks, the bubble price becomes less and less
coupled to the market fundamentals, and the super-exponential
growth of the price makes the market more and more unstable. In this scenario,
the end of the bubble conditional on the absence of crash occurs
at a critical time at which the market becomes maximally unstable.
The end of the bubble is therefore the time when the crash
is the most probable. With or without a crash, the end of the bubble
signals the end of  the transient super-exponential growth,
and the transition to a different regime, with unspecified characteristics.

Here, we assume that sophisticated market participants are indeed aware of the
current  bubble state, and that they know the price is growing towards its final
singularity which will occur at some future random critical time 
at which the market may collapse with a finite probability (but not with certainty).
Our bounded rational agents are able to
form unbiased rational expectations of the critical
time corresponding to the end of the bubble at which the crash 
is the more probable. We assume that 
these sophisticated arbitrageurs enter sequentially into 
the market, attracted by the potential large gains,
given their anticipation of the crash risk quantified by their estimation
of the critical end time of the bubble which is formed when they
enter the market. Because their anticipations of the bubble demise 
are heterogeneous, they solve an optimal timing problem
with distinct inputs, which leads to 
different exit strategies. The heterogeneity
in their exit strategies is common
knowledge among these arbitrageurs, and results in a lack
of coordination, ensuring the persistence of the bubble.
This {\em synchronization problem} is analogous to that identified
by \citet{abreu2003}. However, for
Abreu and Brunnermeier,
the lack of synchronization stems from 
the existence of heterogeneous beliefs on the start of
the bubble, i.e., arbitrageurs have ``sequential awareness''
and do not know whether they have learn the information
on the mispricing early or late relative to other rational 
arbitrageurs. In contrast, our model emphasizes that
the lack of synchronization results from the heterogeneous beliefs 
concerning the end of the bubble. 
Many reports both in the academic and professional literature
state that sophisticated participants like hedge-funds correctly
diagnosed the presence of a bubble and actually ``surfed''
the bubbles, attracted by the potential large gains. 
Many reported that the largest uncertainty was how long it would continue
its course \citep{Gurkaynak2008, Sullivanbubble}

Less sophisticated traders investing in the market
have little knowledge on the bubble duration and their action
add noise which is assumed to have an influence only 
on the critical time characterizing the end of the bubble,
while the super-exponential growth of the price remains robust.
In the bubble regime, a well-defined nonlinear exponent characterizes
the positive feedbacks at the origin of the bubble. The noisy character
of the critical time $t_c$ of the end of the bubble is modeled by 
an Ornstein-Uhlenbeck process. Intuitively, the price trail resembles 
the trace of a bug climbing erratically along
a hanging curved rope attached to a vibrating support. 

Specifically, the price dynamics in the bubble regime is assumed to be described  by the following stochastic differential equation:
\begin{equation}\label{E: stc1}
\df p=\mu p^m (1+\delta(p,t))\df t+\sigma p^m \df W~,
\end{equation}
where the exponent $m>1$ embodies the positive feedback mechanism, in which a high price $p$
pushes even further the demand so that the return and its volatility tend to be a nonlinear accelerating
function of $p$. When $m>1$, we will show that the price diverges in finite time. The 
time at which this divergence occurs is referred to as the critical time $\tc$.
As we will see later, the term $\delta(x,t)$ is a time-varying regulator term that governs the behavior of $\tc$, $\mu$ is the instantaneous return rate, $\sigma$ is the volatility of the returns and $W$ is the standard Wiener process. This model recovers the standard Black-Scholes model of the geometric random walk with drift $\mu$ and standard deviation $\sigma$ for $m=1$ and $\delta=0$.

Let us consider first the case where $\delta(p,t)= \sigma =0$, so that expression (\ref{E: stc1}) reduces to 
$\df p=\mu p^m \df t$, whose solution is
\begin{equation}
p=K(t_c-t)^{-\beta}~,
\label{thrbhwgwg}
\end{equation}
where $\beta=\frac{1}{m-1}$ , $K=(\frac{\beta}{\mu})^{\beta}$, $t_c=\frac{p_0^{-(m-1)}}{(m-1)\mu}$ and $p_0$ denotes the price at the start time of the bubble taken to be $t=0$. Since $\beta >0$ for $m>1$, expression
(\ref{thrbhwgwg}) exemplifies the existence of a finite-time (or ``movable'' , \citet{BenderOrzag}) singularity of the price
that goes to infinity in finite time as $t \to t_c^-$. This pathological behavior is the direct consequence
of the positive feedback embodied in the condition $m>1$.

Motivated by this simple analytical solution, we now consider the more general process (\ref{E: stc1}) 
and specify $\delta(p,t)$ in order to obtain a general process with stochastic finite time singularities.
We postulate the following specific form for the process governing $\delta(p,t)$,
\begin{equation}
\delta(p,t)=\alpha~ \tc(t)+\frac{1}{2}m\mu \sigma^2  ~[p(t)]^{m-1}~,
\label{thykkklgw}
\end{equation}
where
\begin{equation}\label{OU}
\df \tc=-\alpha \tc \df t+(\sigma /\mu) \df W 
\end{equation}
follows an Ornstein-Uhlenbeck process with zero unconditional mean. 
The Wiener process in (\ref{OU}) is the same as the one in (\ref{E: stc1}).
We obtain the following result.
\begin{proposition}\label{l:1}
Provided that $\delta(p,t)$ follows the process (\ref{thykkklgw}) with (\ref{OU}), the solution 
of equation \eqref{E: stc1} can be written under a form similar to (\ref{thrbhwgwg}) as follows,
\begin{equation}\label{6}
p(t)=K(\widetilde{T_c}-t)^{-\beta}~, 
\end{equation}
with
\begin{equation}
\beta=\frac{1}{m-1}~, \quad
K=\left(\frac{\beta}{\mu}\right)^{\beta}~, \quad T_c=\frac{\beta}{\mu} p_0^{-\frac{1}{\beta}}~, \quad
\widetilde{T_c}=T_c+\tc~.
\label{thytnkgmw}
\end{equation}
\end{proposition}
The proof of Proposition \eqref{l:1} is given in Appendix A. 

Note that 
\begin{equation}
\E(\widetilde{T}_c)=\E(T_c+\tc)=T_c~.
\label{hwfca}
\end{equation}
Therefore, $T_c$ given in (\ref{thytnkgmw}) is the expected time at which the bubble
will end in an explosive singularity. The Ornstein-Uhlenbeck process (\ref{OU}) for $\tc(t)$
expresses that the end of the bubble cannot be known with certainty but is instead a
stochastic variable. Nonetheless, due to the positive feedback, the price can explode
in finite time, but the end of the bubble can at best be known to follow the mean-reversing
process  (\ref{OU}). The time $T_c$ can be interpreted as the consensus rational expectation 
formed by sophisticated arbitrageurs of the stochastic critical time
$\widetilde{T_c}$ as which the bubble is expected to end. But each
trading day $t$ discloses a different `actual' critical time $T_c+t_c(t)$ which is vibrating around
its expected value $T_c$. 

Our model assumes that there is no coordination mechanism
that would ensure the exchange of information among the sophisticated arbitrageurs
concerning their expectation of the end of the bubble. The arbitrageurs reveal
their private information only upon entering the market. We assume that they do so 
sequentially, based on their heterogeneous beliefs on the process $\tc$. And
the process $\tc$ is an emergent property that results 
from the aggregation of market beliefs rather than from the action of single arbitrageur. 
Being aware of the escalating level of the bubble that has not yet burst, 
each arbitrageur will ride the bubble for a while and identify the best exit strategy
according to the maximization of her risk-adjusted return based on her belief.

Let us denote $t_i>0$ the time at which the 
$i$'s arbitrageur has entered the market. Being aware of the form (\ref{6}) of the price
dynamics, at each instant $t$, the rational arbitrageur forms a belief quantified by
her hazard rate $h_i(t)$, of
the probability that a crash might occur in the next instant, conditional on the fact
that it has not yet happened. This allows her to estimate the probability $1-\Pi_i(t)$
that the crash will not happen until time $t$. Given the explosive form (\ref{6}) of the price
dynamics, we assume that the arbitrageur forms a belief of the crash hazard rate
which is of the same form, that is,
\begin{equation}\label{E: h}
h_i(t)=\frac{\pi_i(t)}{1-\Pi_i(t)}\propto (T_{c,i}-t)^{-\beta_i}~,
\end{equation}
where $\Pi_i(t)$ is the arbitrageur $i$'s conditional cumulative distribution function of the bursting date and $\pi_i(t)$ represents the associated conditional density. $T_{c,i}=T_c+\tilde{t}_{c,i}$ denotes the critical time 
for the end of the bubble that the arbitrageur $i$ has estimated when entering the market. 
We allow the exponent $\beta_i$ to be different from arbitrageur to arbitrageur, so as to reflect
different views on the riskiness of the market, which can translate into distinct risk aversions:
the larger the exponent $\beta_i$, the more pessimistic is the view of the arbitrageur concerning
the imminence of the crash, because a larger $\beta_i$ implies a faster divergence of the crash
hazard rate.

The occurrence of the market collapse is posited to be triggered when 
a sufficiently large number $\eta$ of arbitrageurs have exited the market, leading to 
a large price movement, amplified by the herding of noise traders. Their cumulative effect
is accounted for by a postulated percent loss $\kappa$ of the crash, which is
itself a random variable.
Given such an environment, the date to exit the market for a given rational arbitrageur $i$  
determines her best trade strategy, which is found as the solution of the 
following optimization problem
 \begin{equation}\label{E: opt}
\max_{t}\,\, \E^i[(1-\Pi_i(t))\cdot \df p-\pi_i(t)\df t\cdot \kappa p]~.
\end{equation}
The first term $(1-\Pi_i(t))\cdot \df p$ represents the arbitrageur's instantaneous benefit at $t+dt$ provided that the burst of the bubble has not yet happened. The second term $\pi_i(t)\df t\cdot \kappa p$ is the instantaneous cost supported by the arbitrageur  when the bubble bursts. The solution of \eqref{E: opt} is obtained from the first-order condition 
\begin{equation}
(1-\Pi_i(t))\E^i(\mu p^{m}(1+\delta(p,t)))=\pi_i(t) \E^i(\kappa p)~.
\end{equation}
We can thus state
\begin{proposition}\label{proptimeexit}
Given a population of heterogeneous arbitrageurs, which form their expectation
of the crash hazard rate according to (\ref{E: h}) with heterogeneous
anticipated critical times $T_{c,i}$ and exponents $\beta_i$ reflecting their
different views on the riskiness of the market, a given arbitrageur $i$ decides
to exit the market at the date $t_i^{ex}$ which is the solution of 
\begin{equation}
{\E^i[dp(t^{ex}_i)] \over \E^i[\kappa p(t^{ex}_i)]} =h_i(t^{ex}_i)\propto (T_{i,c}-t^{ex}_i)^{-\beta_i}~.
\label{thj2otjg2tpg}
\end{equation}
Since $\E^i(p)\neq\E^j(p)$ and $h_i(t)\neq h_j(t)$, we have $t_i^{ex}\neq t_j^{ex}$. Notwithstanding the fact that the presence of the bubble is common knowledge among all rational arbitrageurs, the absence of synchronization 
of their market exit allows the bubble to persist and run its course up to a time close to its expected value (\ref{hwfca}).
\end{proposition}
For the price process (\ref{E: stc1}) with (\ref{thykkklgw}), equation (\ref{thj2otjg2tpg}) yields
\begin{equation}
\frac{\E^i(\mu p^m(1+\alpha \tilde{t}_c(t^{ex}_i)+\frac{1}{2\mu}m \sigma^2 [p(t^{ex}_i)]^{m-1}))}{\E^i(\kappa p)}=h_i(t^{ex}_i)\propto (T_{i,c}-t^{ex}_i)^{-\beta_i}~.
\end{equation}
This {\em synchronization problem} is analogous to that identified
by \citet{abreu2003}, with the important difference that we emphasize that
the lack of synchronization results from the heterogeneous beliefs 
concerning the end of the bubble.

The corresponding observable logarithmical return for the asset price reads
\begin{align}
r_\tau(t)=\ln p(t+\tau)-\ln p(t)&=-\beta\ln\left(\frac{T_c+\tilde{t}_{c}(t+\tau)-{(t+\tau)}}{T_c+\tilde{t}_{c}(t)-{t}}\right) \nonumber \\
&=-\beta\ln\left(1+\frac{\Delta \tilde{t}_{c}(t)-\tau}{T_c+\tilde{t}_{c}(t)-t}\right)~,
\label{E: 17}
\end{align}
where $\tau$ is the time interval between two observations of the price. 
In the case where $T_c$ is large enough such that $T_c+\tilde{t}_{c}(t)-t\gg \Delta\tilde{t}_{c}(t)-\tau$, 
expression \eqref{E: 17} can be approximated by its first order Taylor expansion:
\begin{equation}
r_\tau(t)=\frac{\beta}{T_c+\tilde{t}_{c}(t)-t}(\tau-\Delta \tilde{t}_{c}(t))=\frac{1}{(m-1)(T_c+\tilde{t}_{c}(t)-t)}(\tau-\Delta \tilde{t}_{c}(t))\label{E: 18}
\end{equation}
Therefore, under the condition $m\to1$ and $T_c+\tilde{t}_{c}(t)-t \to \infty$, the logarithmical return $r_\tau(t)$ is driven by  the change $\Delta \tilde{t}_{c}(t)$ of the critical time on each trading day. 
Although $\tc$ follows an Ornstein-Uhlenbeck process, i.e., $\Delta \tilde{t}_{c}(t)=-\alpha  \tilde{t}_{c}(t)+\varepsilon_t$, 
the existence of correlation between successive returns will be hardly detectable if $\alpha$ is small enough. Conversely, if $\tc$ follows a unit root process, the logarithmical return $r_\tau(t)$  is only dependent on the 
realization of the gaussian noise term $\varepsilon$. In this sense, only if $T_c$ is not too far in the future,  $m$ is sufficiently larger than $1$ and $\tc$ is stationary, can we diagnose the existence of a bubble, 
characterized by the emergence of a transient super-exponential price growth of the form (\ref{6}).

\section{Second bubble model: finite-time singularity in the momentum price dynamics with stochastic critical time
\label{th3wacvqtegbtt4}}

The price process (\ref{6}) of the first bubble model might appear extreme, in the sense that 
the price diverges on the approach of the critical time $\widetilde{T_c}$ of the end of the bubble.
However, such a divergence cannot run its full course in our model due to the divergence of the crash hazard rate
which ensures that a crash will occur before. The critical time is thus a ghost-like time, which 
is out-of-reach, and the price process (\ref{6}) describes a transient run-up that would diverge
only in the hypothetical absence of any arbitrageur. Here, we consider an alternative model in which the price
remains always finite but the faster-than-exponential growth associated with the bubble is
embodied into the price momentum, i.e., the derivative of the logarithm of the price or logarithmic return.

Defining $y(t) = \ln p(t)$, we assume the following process for $y(t)$:
\begin{align}
\df y&=x(1+\gamma(x,t))\df t +(\sigma / \mu) x \df W   \label{stcII part1}\\
\df x&= \mu x^m(1+\delta(x,t))\df t+\sigma x^m\df W~,
\label{stcII part2}
\end{align}
where the same Wiener process $W(t)$ acts on both $\df y$ and $\df x$.
The process $x(t)$ plays the role of an effective price momentum. To see this,
consider the special case $\gamma(x,t)=0$. Then, expression (\ref{stcII part1}) reduces to 
$\df y=x \df t +(\sigma / \mu) x \df W$, which shows that $x(t) \df t =  \E [ \df y]$ and thus $x(t)$
is the average momentum of the price, defined as the instantaneous time derivative
of the expected logarithm of the price. The dynamics of the log-price described by (\ref{stcII part1})
with (\ref{stcII part2}) is similar to previous models 
\citep{Bouchaud-Cont98,Farmer98,IdeSornette1,IdeSornette2}, which argued for the
presence of some inertia in the price formation process. This inertia is related to the momentum effect \citep{jegadeesh1993,jegadeesh2001,Carhart97,Xue03,Cooperetal04}.
Intuitively, a price process involving both $\df y$ and $\df x$ holds if
the price variation from today to tomorrow is based in part on 
decisions using analyses of the price change between yesterday (and possibly earlier times) and today.

In the (unrealistic) deterministic limit  $\gamma(x,t)=\delta(x,t)=\sigma=0$, the two
equations (\ref{stcII part1}) and (\ref{stcII part2}) reduce to the deterministic equation
\begin{equation}\label{E: inertia}
\frac{\df^2 y}{\df t^2}=\mu \left(\frac{\df y}{\df t}\right)^m ~,
\end{equation}
whose solution reads, for $m>2$,
\begin{equation}
y(t)=A-B(T_c-t)^{1-\beta}~,
\label{thjapfvmqefbvqe}
\end{equation}
where $\beta=\frac{1}{m-1}$, $T_c=(\beta /\mu) (\frac{\df p}{\df t}\big|_{t=t_0})^{-\frac{1}{\beta}}$, 
$B=\frac{1}{1-\beta}(\mu / \beta)^{-\beta}$ and $A=p(T_c)$. 
The condition $m>2$ ensures that $0 < 1-\beta <1$. Therefore, the log-price $y(t)$
exhibits a finite-time singularity (FTS) at $T_c$. But this FTS is of a different type than
in the model of the previous section: here, $y(t)$ remains finite at $t=T_c$
and equal to some value $A=p(T_c)$. The singularity is expressed via the divergence
of the momentum $x(t) = dy/dt$ which diverges at $t=T_c$. As in the previous model,
this FTS embodies the positive feedback mechanism, in which a high price momentum $x$
pushes even further the demand so that the return and its volatility tend to be nonlinear accelerating
functions of $x$. In the previous model, it is the price that provides a feedback on further price 
moves, rather than the price momentum used here.

Motivated by this simple analytical solution (\ref{thjapfvmqefbvqe}), 
we complement the general process (\ref{stcII part1},\ref{stcII part2}) 
by specifying $\gamma(x,t)$ and $\delta(p,t)$ 
in order to obtain solutions with stochastic finite time singularities in the 
momentum with finite prices.
We postulate the following specific processes 
\begin{align}
\gamma(x,t)&=\alpha\tc(t)+\frac{\sigma^2}{2\mu} [x(t)]^{m-1}  \label{hwinvdwdv} \\
\delta(x,t)&=\alpha\tc(t)+\frac{1}{2}m\mu \sigma^2 [x(t)]^{m-1}  \label{hwinvdwdv2}
\end{align}
where
\begin{equation}\label{OU2}
\df \tc=-\alpha \tc \df t+(\sigma /\mu)\df W 
\end{equation}
follows an Ornstein-Uhlenbeck process with zero unconditional mean. 
The Wiener process in (\ref{OU2}) is the same as the one in (\ref{stcII part1},\ref{stcII part2}),
which reflects that the same series of news or shocks move log-price, momentum 
and anticipated critical time. We obtain the following result.
\begin{proposition}
\label{thjbjew}
Provided that  $\gamma(x,t)$ and $\delta(p,t)$ follow 
the processes given respectively by (\ref{hwinvdwdv}) and
 (\ref{hwinvdwdv2}), then the solution of (\ref{stcII part1},\ref{stcII part2}) for
 the log-price $y(t)=\ln p(t)$ can be written under a form similar to expression (\ref{thjapfvmqefbvqe}) as follows,
\begin{equation}\label{E: stcII}
y(t) = A-B(T_c+\tc(t)-t)^{1-\beta}
\end{equation}
where 
\begin{equation}
\beta=\frac{1}{m-1}~, \quad  T_c=\frac{\beta}{\mu} x_0^{1/\beta}, ~ \quad x_0 := x(t=0)~, \quad
B=\frac{1}{1-\beta}(\beta / \mu)^{\beta}~,
\label{hhmojun}
\end{equation}
and $A$ is a constant.
\end{proposition}
The proof of Proposition \ref{thjbjew} is given in appendix B. 

Expression (\ref{E: stcII})
describes a log-price trajectory exhibiting a FTS occurring at an unknown future critical
time $T_c+\tc(t)$ which itself follows an Ornstein-Uhlenbeck walk.
In the same manner as in the previous model of the preceding section, 
we assume that there is no coordination mechanism
that would ensure the exchange of information among the sophisticated arbitrageurs
concerning their expectation of the time $T_c+\tc(t)$  of the end of the bubble. 
Following step by step the same reasoning as in the previous section, we 
conclude that  Proposition \ref{proptimeexit} also holds for the present model.
Being aware of the escalating level of the bubble that has not yet burst, 
each arbitrageur will ride the bubble for a while and identify the best exit strategy
according to the maximization of her risk-adjusted return based on her belief.
Proposition 2 determines that the exit time $t^{ex}_i$ for arbitrageur $i$ is the solution of 
\begin{equation}
{\E^i\left[ p(t^{ex}_i) x(t^{ex}_i) (1+\gamma(x(t^{ex}_i),t^{ex}_i)) + {\sigma^2 \over 2\mu^2} p(t^{ex}_i) [x(t^{ex}_i)]^2\right] \over \E^i\left[ \kappa p(t^{ex}_i) \right]} =h_i(t^{ex}_i)\propto (T_{i,c}-t^{ex}_i)^{-\beta_i}~.
\end{equation}

The observable logarithmic return for the asset price corresponding to  \eqref{E: stcII} reads
\begin{align}
r_\tau(t) = y(t+\tau) - y(t) &= -B[(T_c+\tilde{t}_{c}(t+\tau)-(t+\tau))^{1-\beta}-(T_c+\tilde{t}_{c}(t)-t)^{1-\beta}]\\
&=-B (T_c+\tilde{t}_{c}(t)-t)^{1-\beta}[(1+\frac{\Delta \tilde{t}_{c}(t)-\tau}{T_c+\tilde{t}_{c}(t)-t})^{1-\beta}-1]\label{E: 33}
\end{align}
For future potential critical times $T_c$ sufficiently far away from the present time $t$ 
such that $T_c+\tilde{t}_{c}(t)-t\gg \Delta\tilde{t}_{c}(t)-\tau$, expression \eqref{E: 33} can be simplified into
\begin{equation}
r_\tau(t) =-\frac{(1-\beta)B}{(T_c+\tilde{t}_{c}(t)-t)^{\beta}}(\Delta \tilde{t}_{c}(t)-\tau)=\frac{\mu^{-\beta}}{[(m-1)(T_c+\tilde{t}_{c}(t)-t)]^{\beta}}\cdot(\tau-\Delta \tilde{t}_{c}(t))\label{E: 34}
\end{equation}
Eq.\eqref{E: 34} has a structure similar to that of Eq.\eqref{E: 18}. 
For weak positive feedback of the momentum on itself ($m \to 2^+$)
and when $T_c$ is large enough so that $\mu(T_c+\tilde{t}_{c}(t)-t)$ is slowly varying, 
then the logarithmical return $r_s$ is essentially driven by $\Delta \tilde{t}_{c}(t)$,  i.e., 
the change of critical time disclosed by every trading day is the main stochastic process. 
The Geometric Brownian Motion is then recovered as an approximation in this limit when 
the correlation time of the Orstein-Uhlenbeck process driving the critical goes to zero.

\section{Empirical tests of the two bubble models} \label{S: 4}

We have proposed two models in which a financial bubble
is characterized by a transient faster-than-exponential growth culminating into a finite-time singularity
at some potential critical time. Because our two models reduce to a standard GBM in appropriate limits,
the diagnostic of the presence of bubbles according to our two models lies
in the conjunction of three pieces of evidence that characterize
specific deviations from the GBM regime:  (i) the proximity of the calibrated
potential critical date $T_c$ to the end of the time window in which the calibration
of the models are made; (ii) the reconstructed time series of the 
critical time $\tc$ should be stationary and thus reject a standard unit-root test; (iii) the critical exponent $m$ 
should be significant larger than $1$, a condition for the existence
of the super-exponential regime proposed to characterize bubbles.

\subsection{Construction of alarms from the first model}\label{S: 4.1}

Given a financial time series of close prices at the daily scale, our purpose is to develop a procedure using
the model of section \ref{ththwrbw} to diagnose the presence of bubbles.
We use time windows of 750 trading days that we slide with a time step of 25 days
from the beginning to the end of the available financial time series. 
The number of such windows is therefore equal to the total
number of trading days in the financial time series minus 750 and divided by 25.
For each window, the purpose is to decide if the model of section \ref{ththwrbw}
diagnoses an on-going bubble or not and then to compare with the actual 
subsequent realization of a crash that we consider as the validation step.

For each window $]t_i-750, t_i]$ ending at $t_i$, we transform 
the price time series in that window into a critical time series
by inverting expression \eqref{6} for $\widetilde{T}_{c,i}(t)$:
\begin{equation}
\widetilde{T}_{c,i}(t) =\frac{1}{K} {1 \over [p(t)]^{1/\beta}} +t, \qquad t=t_i-749,\cdot\cdot\cdot,t_i~.
\label{yhyhiwpw}
\end{equation}
The critical time series $\widetilde{T}_{c,i}(t)$ is defined over the window $i$ ending at $t_i$.
If the model was exact and no stochastic component was present, and in absence of 
estimation errors, $\widetilde{T}_{c,i}(t)$ would be a constant equal to $T_c$
defined in (\ref{thytnkgmw}).
In the presence of an expected strong stochastic component, we estimate $T_c$
according to (\ref{hwfca}) as
the arithmetical average of $\widetilde{T}_{c,i}(t)$ 
\begin{equation}
T_{c,i}=\frac{1}{750}\sum_{t=1}^{750}\widetilde{T}_{c,i}(t)~.
\label{yhnmbmn}
\end{equation}
We can then construct $\tilde{t}_{c,i}(t)$ as
\begin{equation}
\tilde{t}_{c,i}(t)=\widetilde{T}_{c,i}(t)-T_{c,i}~.
\label{tyhjjvwow}
\end{equation}
The transformation (\ref{yhyhiwpw}) from a non-stationary possibly explosive
price process $p(t)$ into what should be a stationary time series $\widetilde{T}_{c,i}(t)$
in absence of misspecification is a key element of our methodology
for bubble detection that avoids the problems documented by  
 \cite{Granger1974} and  \cite{Phillips1986}
resulting from direct calibration of price or log-price time series.

It will not have escaped the attention of the reader that the transformation (\ref{yhyhiwpw})
requires the knowledge of the two unknown parameters $K$ and $\beta$ that 
specify the bubble process \eqref{6}. We propose to determine these two parameters
by applying an optimization procedure as follows. Recall that a crucial ingredient of the bubble
model is the mean reversal nature of the potential critical time $\tilde{t}_{c}$. This 
suggests to apply a unit root test on the reconstructed time series $\tilde{t}_{c,i}(t)$
and determine the optimal values $K_i^*$ and $\beta_i^*$ as those which 
make the time series $\tilde{t}_{c,i}(t)$ as stationary as possible. We proceed in two steps.
We first search in the space of the two parameters $K$ and $\beta$
and select an elite list of the ten best pairs $(K, \beta)$ (when they exist) which reject a standard unit-root test of
non-stationarity at the 99.5\% significance level. We implement this procedure
with the t-test statistic  of the Dickey-Fuller unit-root test (without intercept).
Since the Dickey-Fuller test is a lower test, the smaller the statistics $t$, 
the larger is the probability to reject the null hypothesis that $\tc$ has unit-root (is non-stationary). 
Of course, only a subset of the windows will yield any solution at all, i.e., 
it is quite often the case that the Dickey-Fuller unit-root test is not rejected at 
the 99.5\% significance level for any pair $(K, \beta)$. For those windows for 
which there are selected pairs $(K, \beta)$ according to the Dickey-Fuller test,
we choose the one with the smallest 
variance for its corresponding time series $\tilde{t}_{c,i}(t)$, i.e., such that the
$\tilde{t}_{c,i}(t)$'s are the closest to their mean in the variance sense. This yields
the optimal $K_i^*$ and $\beta_i^*$ that best ``fit'' the window $i$ in the sense
that this pair of parameters provides the closest approximation to a stationary  time series for
$\tilde{t}_{c,i}(t)$ given by expression (\ref{tyhjjvwow}) for the potential termination of the bubble.

\begin{figure}[h]
\centering
\includegraphics[width=18cm]{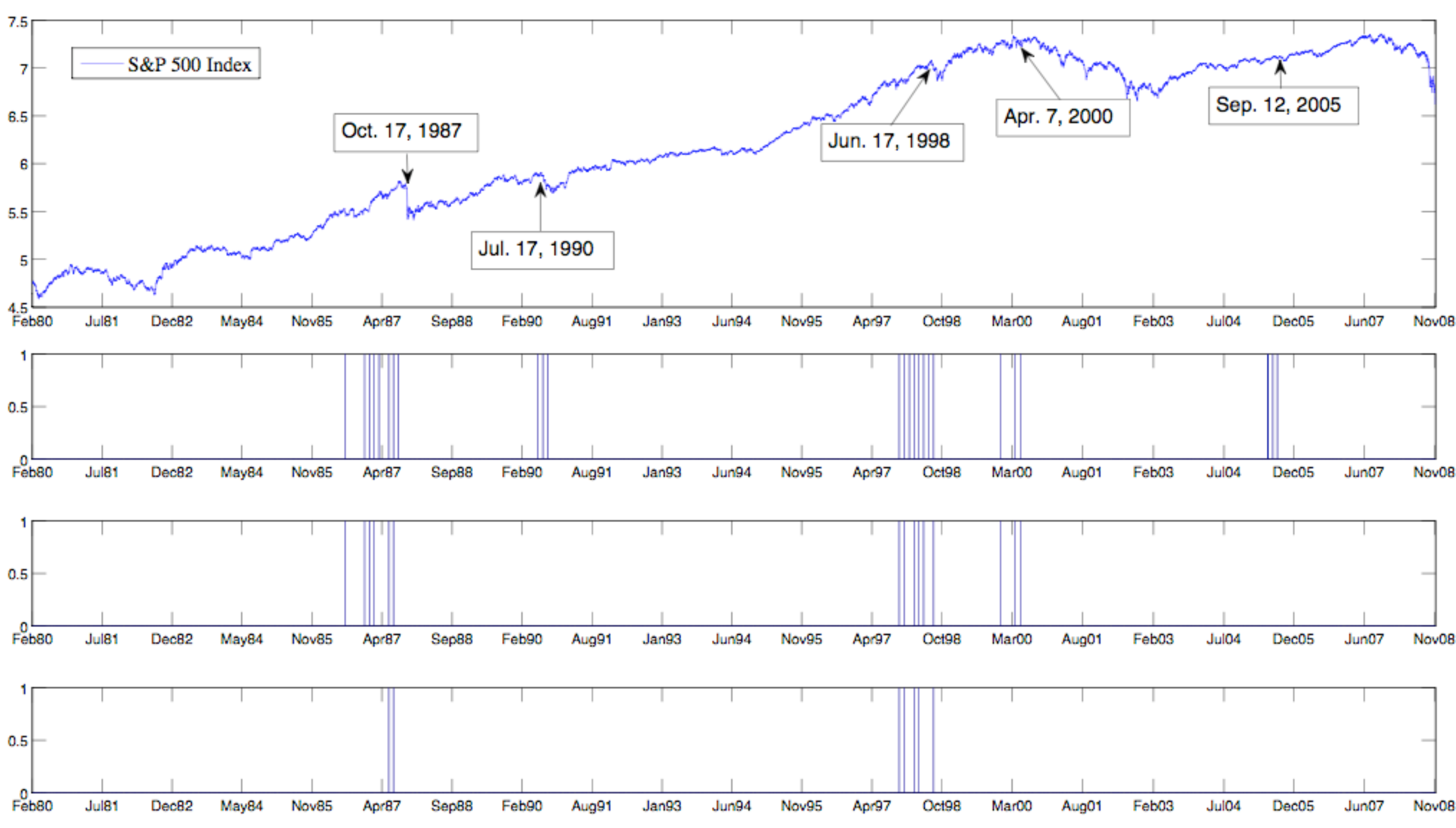}
\caption{Logarithm of the historical S\&P500 stock index and corresponding alarms shown
in the three lower panels as vertical lines indicating the ends of the windows of 750 trading days
in which our procedure using the first bubble model of section 2 flags a diagnostic for the presence of bubble. The three
lower panels corresponds to alarms for which
$T_{c,i} -t_i < 750$, $T_{c,i} -t_i < 500$ and $T_{c,i} -t_i < 250$, from top to bottom.
By definition, the set of alarms of the lowest panel is included in the set of alarms of the middle panel
which is itself included in the set of alarms of the upper panel.
The exponents $m$ found for the upper panel corresponding to $T_{c,i} -t_i < 750$
have a mean of $2.76$ with a standard deviation of $0.33$.}\label{F: 1}
\end{figure}

\begin{figure}[h]
\centering
\includegraphics[width=18cm]{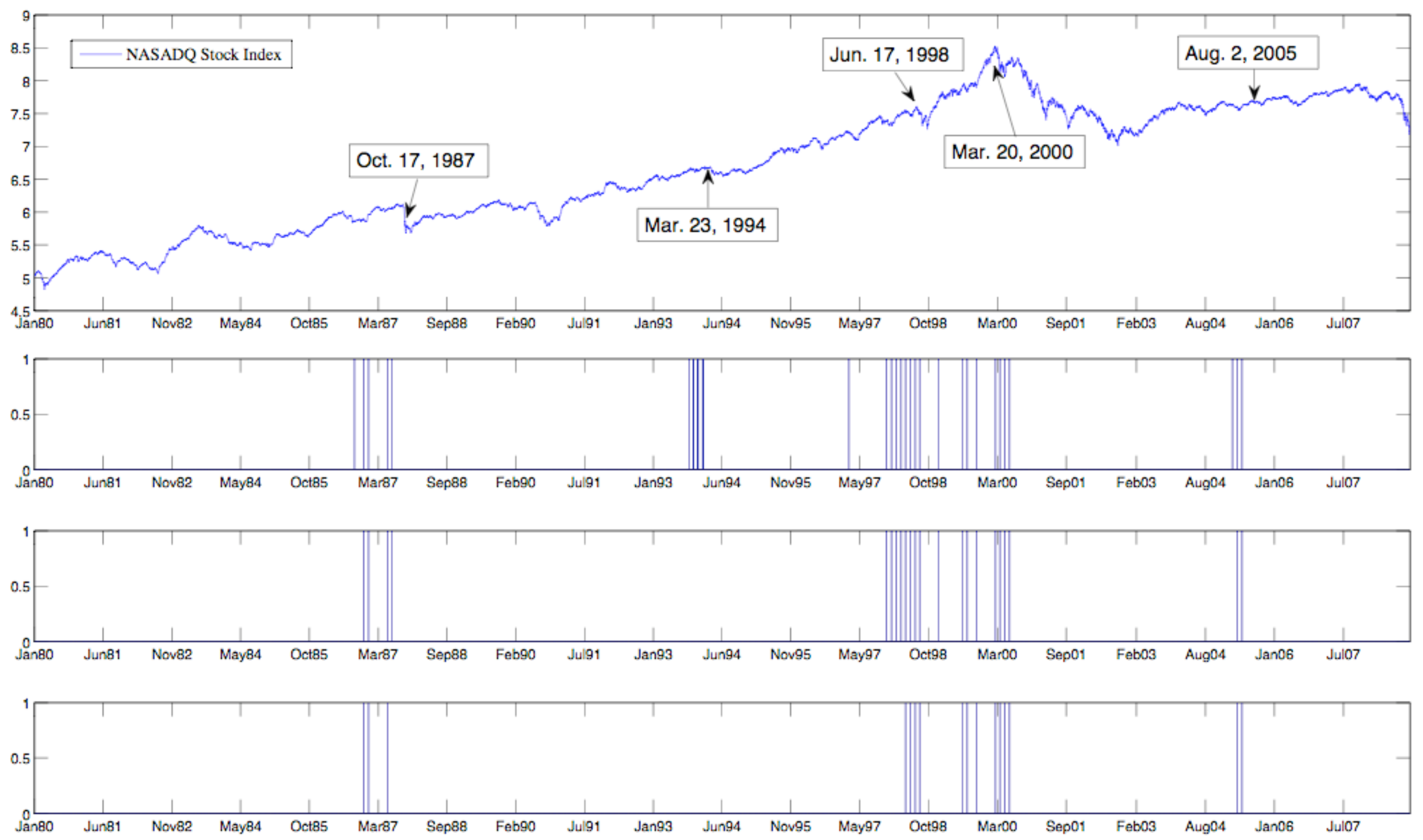}
\caption{Same as Fig.\protect\ref{F: 1} for the Nasdaq Composite index. The exponents $m$ found for the upper panel corresponding to $T_{c,i} -t_i < 750$ have a mean of $2.85$ with a standard deviation of $0.23$.}\label{F: 2}
\end{figure}

\begin{figure}[h]
\centering
\includegraphics[width=18cm]{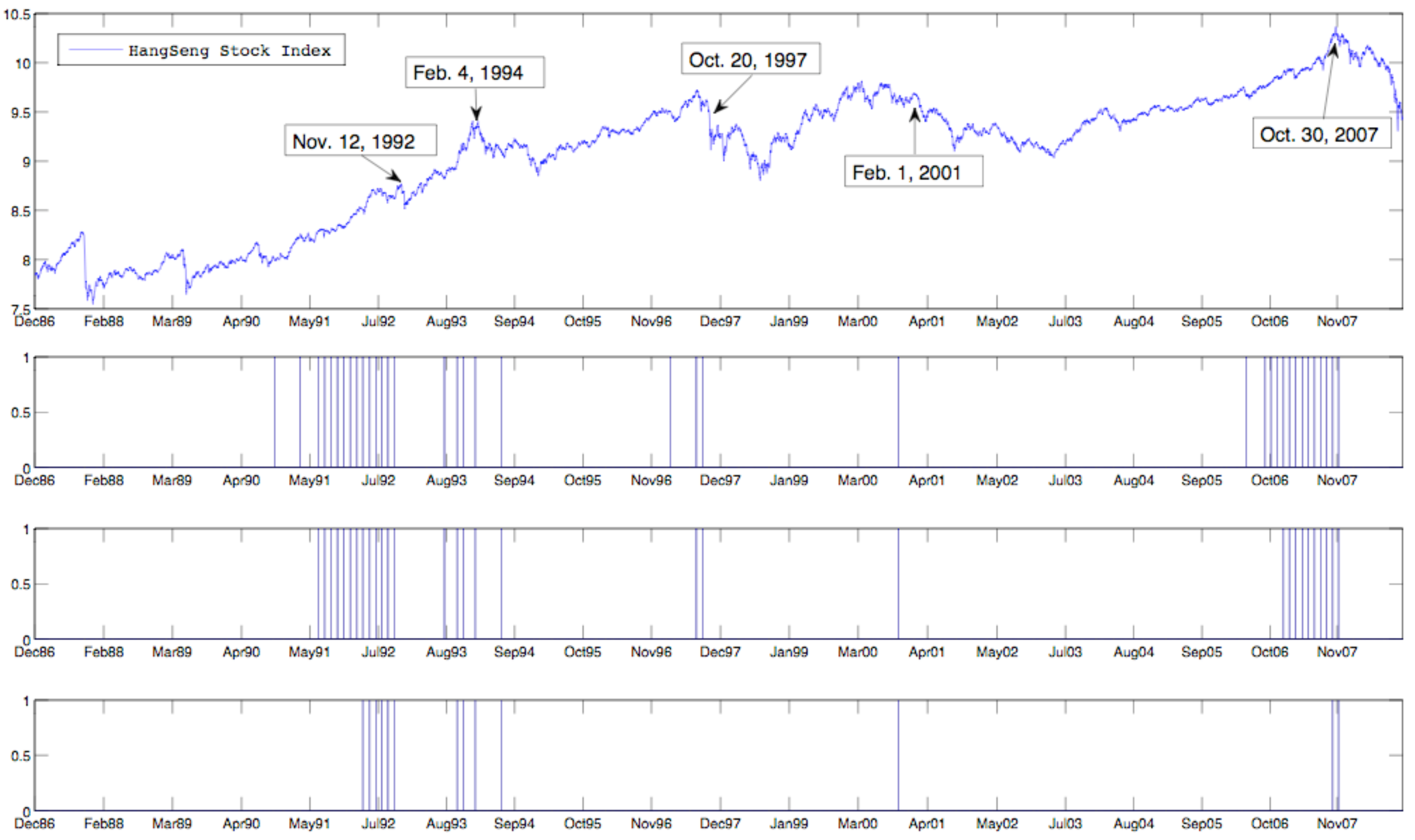}
\caption{Same as Fig.\protect\ref{F: 1} for the Heng Seng index of Hong Kong. The exponents $m$ found for the upper panel corresponding to $T_{c,i} -t_i < 750$ have a mean of $2.84$ with a standard deviation of $0.22$.}\label{F: 3}
\end{figure}

For a given window $i$, a diagnostic for the presence of bubble is flagged and an alarm is declared
when 
\begin{enumerate}
\item[(i)] $\beta^*>0$ such that $m>1$ (the signature of a positive feedback in our model) and
\item[(ii)] $T_{c,i} -t_i < 750$, i.e., the estimated termination time of the bubble is not too distant.
\end{enumerate}
Figs.\ref{F: 1}-\ref{F: 3} depicts all the bubble alarms obtained by applying this procedure
to three major stock indices, the US S\&P500 index from 1 February 1980 to 31 October 2008, 
the US NASDAQ composite index from 1 January 1980 to 31 July 2008 and the
Hong Kong Hang Seng index from 1 December 1986 to 30 November 2008.
An alarm is indicated by a vertical line positioned on the last day $t_i$ of the
corresponding window that passes the two criteria (i-ii). We refine the diagnostic
by presenting three alarm levels, corresponding respectively to 
$T_{c,i} -t_i < 750$, $T_{c,i} -t_i < 500$ and $T_{c,i} -t_i < 250$: the closer
the estimated termination of the bubble, the stronger should be the evidence for the bubble
as a faster-than-exponential growth. Another indication is the existence of clustering
of the alarms. If indeed a bubble is developing, it should be diagnosed repeatedly by
several successive windows. 

Fig.\ref{F: 1} for the S\&P500 index shows that the alarm clusters correctly identify the upcoming of four 
significant market corrections or crashes. This suggests to qualify these alarms
as correct diagnostic of bubbles ending in these corrections.
The first one is the large cluster localized within 1.5 years
before the `Black Monday' crash of October 1987. The second smaller cluster is associated
with a significant correction starting in July 17, 1990. The third large cluster is also within
1.5 year before the occurrence of the turmoil starting in August 1998, associated with the
default of Russia on its debt and the devaluation of the ruble. The detection of a bubble
starting to develop more than a year before this event suggests that 
this event may not have been entirely
exogenous, supporting previous evidence for this claim \citep{Whycrash}. The fourth smaller cluster announced
the turning point of the famous Internet-Communication-Technology (ICT) bubble in April 2000.
We note that the strength of the bubble is quite weak for the S\&P500 index. This can probably be
explained by the fact that
only about 20\% of its constituting firms belonged to the ICT sector while
the remaining 80\% firms belonged to the ``old economy'' sector. In contrast, the alarm
signal is much stronger for the Nasdaq index, as can be seen in figure \ref{F: 2}.
One can observe a fifth rather small cluster of alarms for $T_{c,i} -t_i < 750$ (which however disappears
for $T_{c,i} -t_i < 500$ or smaller, suggesting a weak signal) which is dated Sept. 12, 2005 in the top panel 
of Fig.\ref{F: 1}. This alarm does not appear to be associated to any nearby termination of 
a rising price regime. However, notice that, in October 2007, the S\&P500 peaked
and then started a dramatic accelerating downward spiral
fueled by the unfolding of the global financial crisis. This peak of October 2007 is indeed
less than 750 trading days away from the triggering of the alarm on Sept. 12, 2005.
This supports other evidence that the run-up of the S\&P500 from 2003 to October 2007
was a bubble \citep{SorWood09}. It is however a failure of the present methodology
that the alarm is short-lived and does not confirm the continuing accelerating trend up
to the peak in October 2007.

Fig. \ref{F: 2} paints a similar picture. A first bubble preceding the crash of October 1987 is 
clearly diagnosed. A very large cluster of alarms spans the period from
at least early 1997 \citep{phillips2007}  to 2000, confirming the diagnostic of a running ICT bubble, 
that ended with a crash in April 2000. This large cluster is actually made of two
sub-clusters, the first one associated with the bubble behavior ending with the so-called Russian crisis
at the end the summer of 1998, and the second one corresponding to the 
well-known ICT bubble reflecting over-optimistic expectation of a ``new economy''. This is similar
to the analysis and conclusion obtained in Fig.\ref{F: 1} for the S\&P500 index.
There is small cluster of alarms ending in April 1994,
which cannot be associated with any large price movement afterwards.
Finally, a fourth cluster of alarms is dated Sept. 12, 2005 in the top panel 
of Fig.\ref{F: 2}. As for the S\&P500 index,
this cluster of alarms does not appear to be associated to any nearby termination of 
a rising price regime, but rather to the development of an accelerating upward trending
price that culminated in October 2007 before crashing in the subsequent year.

Similar conclusions hold for the Hong Kong market, as shown in Fig.\ref{F: 3}.
One can observe the clusters of alarms associated with the successive booming phases of the Hang Seng index
followed by several corrections or crashes. One can identify in particular the bubbles
associated with the strong correction of 1992, the two Asian crises of 1994 and 1997, as well as
the bubble ending in October 2007, which is this time very clearly diagnosed for this Hong Kong market.
There is one isolated false alarm dated Feb. 1, 2001 in the top panel 
of Fig.\ref{F: 3}.

\subsection{Construction of alarms from the second model \label{S: 4.2}}

Similarly to the procedure described in section \ref{S: 4.1}, we transform a given price time series
in a given window $i$ of 900 successive trading days
into what should be a stationary time series of potential critical end times, if the price series
is indeed described by the bubble model of section \ref{th3wacvqtegbtt4}.
Inverting expression (\ref{E: stcII}) in Proposition 3, we get, similarly to expression (\ref{yhyhiwpw}),
\begin{equation}
\widetilde{T}_{c,i}(t) = t_i + \left(\frac{A-\ln p(t)}{B}\right)^{\frac{1}{1-\beta}}~, \qquad t=t_i-899,\cdot\cdot\cdot,t_i~.
\label{hnomm5pun}
\end{equation}
The critical time series $\widetilde{T}_{c,i}(t)$ is defined within the window $i$ ending at $t_i$.
If the model was exact and no stochastic component was present, and in absence of 
estimation errors, $\widetilde{T}_{c,i}(t)$ would be a constant equal to $T_c$
defined in (\ref{hhmojun}). The expected critical end time $T_{c,i}$ of a bubble, if any, is then estimated 
for this window $i$ by expression (\ref{yhnmbmn}) (with $750$ replaced by $900$). The fluctuations around $T_{c,i}$ are
described by $\tilde{t}_{c,i}(t)$ defined by (\ref{tyhjjvwow}).

As for the first bubble model, the transformation (\ref{hnomm5pun}) requires the determination
of parameters, here the triplet $(A, B, \beta)$. For this, we proceed exactly as in the previous subsection,
with the Dickey-Fuller unit-root test applied to the time series $\tilde{t}_{c,i}(t)$, followed by the
selection of the best triplet $(A^*, B^*, \beta^*)$ that minimize the variance of the time series $\tilde{t}_{c,i}(t)$.
The search of the additional parameter $A$ is
performed in an interval bounded from above by  $2 \max_{t_i-899 \le t \le t_i} \ln p(t)$.
Then, for a given window $i$, a diagnostic for the presence of bubble is flagged and an alarm is declared
when 
\begin{enumerate}
\item[(i)] $0 <\beta^*<1$ such that $m>2$ (the signature of a positive feedback in the momentum price
dynamics model) and
\item[(ii)] $-25 \leq T_{c,i} -t_i \leq 50$,
such that the estimated termination time of the bubble is close to the right side of the time window.
\item[(iii)] We further refine the filtering by considering three levels of significance quantified
by the value of the exponent $m$: level 1 ($m>2$), level 2 ($m>2.5$) and level 3 ($m>3$).
\end{enumerate}
The condition $T_{c,i} -t_i \leq 50$ is much more stringent that its counterpart for the first bubble model. 
The rational is that the price dynamics in terms of a finite-time singularity in the price momentum
corresponds to a weaker singularity that can only be observed, in the presence of a strong stochastic
component, rather close to the potential singularity. This explains the smaller upper bound of $50$
trading days (corresponding approximately to two calendar months). The lower bound of $-25$ days
accounts for the fact that the analysis is performed in sliding windows with a time step of 
$25$ trading days.

\begin{figure}[h]
\centering
\includegraphics[width=18cm]{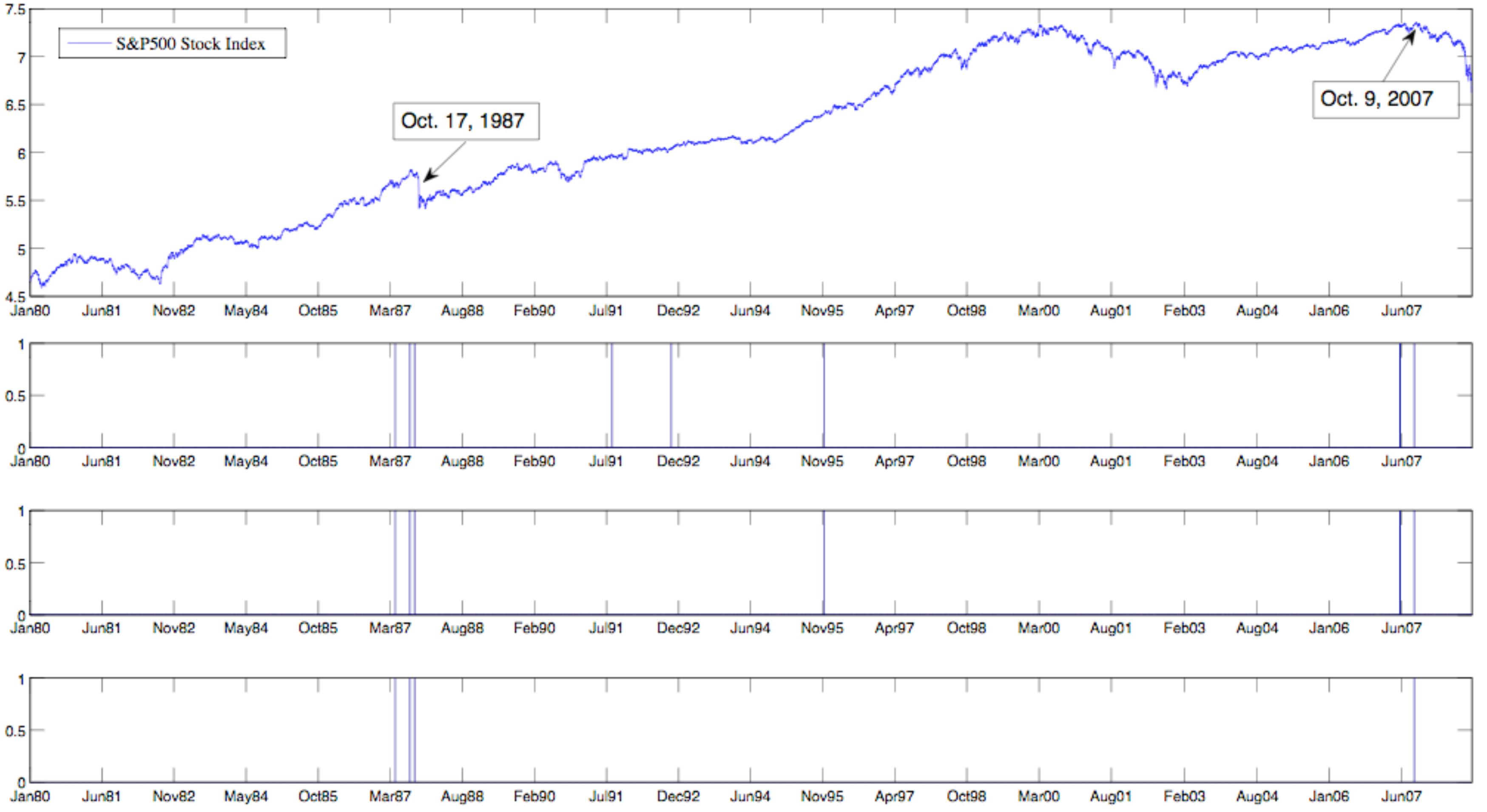}
\caption{Logarithm of the historical S\&P500 stock index and corresponding alarms shown
in the three lower panels as vertical lines indicating the ends of the windows of 900 trading days
in which our procedure using the second bubble model of section 3 flags a diagnostic for the presence of bubble. The three
lower panels corresponds to alarms for which
$m>2$, $m>2.5$ and $m>3$, from top to bottom.
By definition, the set of alarms of the lowest panel is included in the set of alarms of the middle panel
which is itself included in the set of alarms of the upper panel.}\label{F: 4}
\end{figure}

\begin{figure}[h]
\centering
\includegraphics[width=18cm]{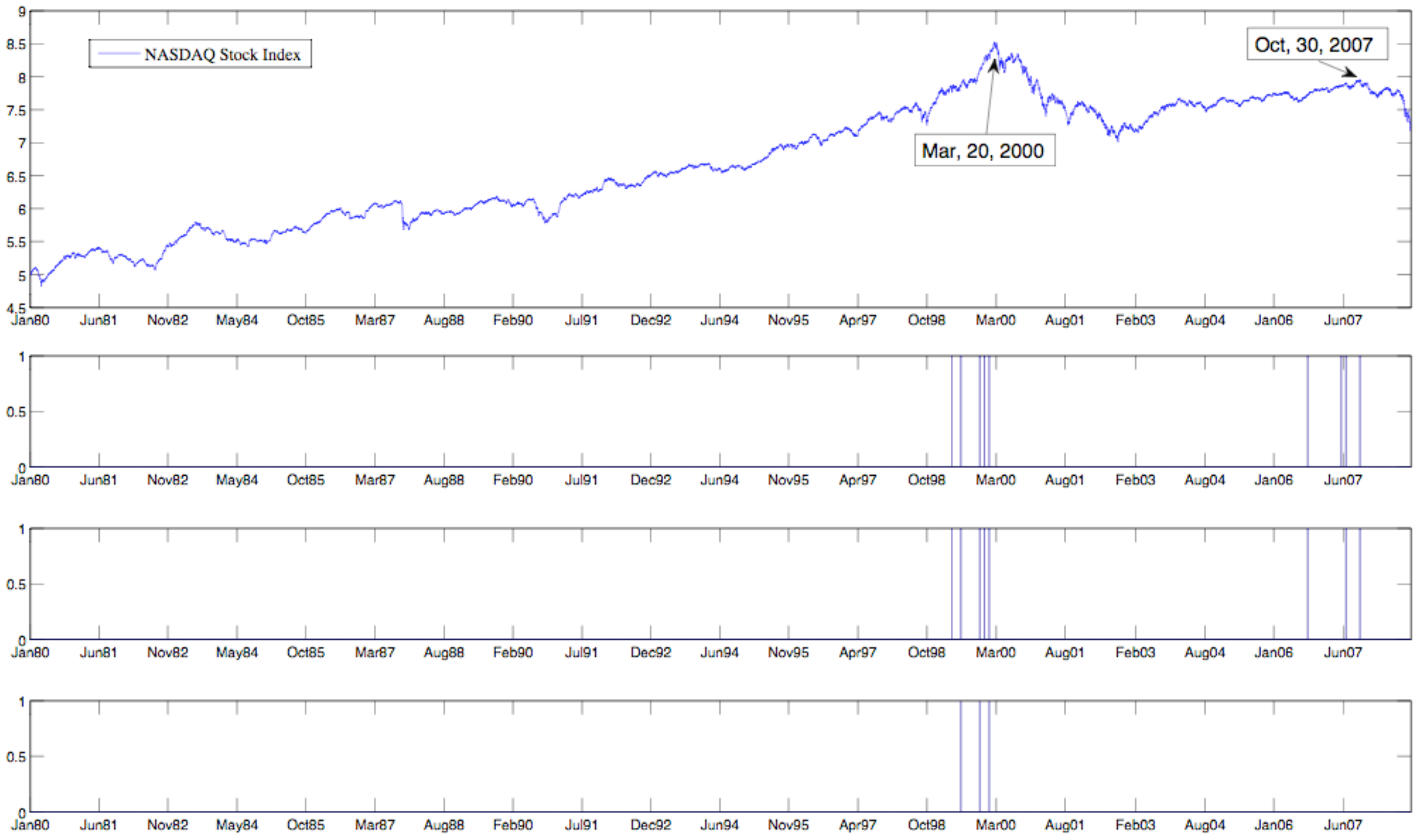}
\caption{Same as Fig.\protect\ref{F: 4} for the Nasdaq Composite index. }\label{F: 5}
\end{figure}

\begin{figure}[h]
\centering
\includegraphics[width=18cm]{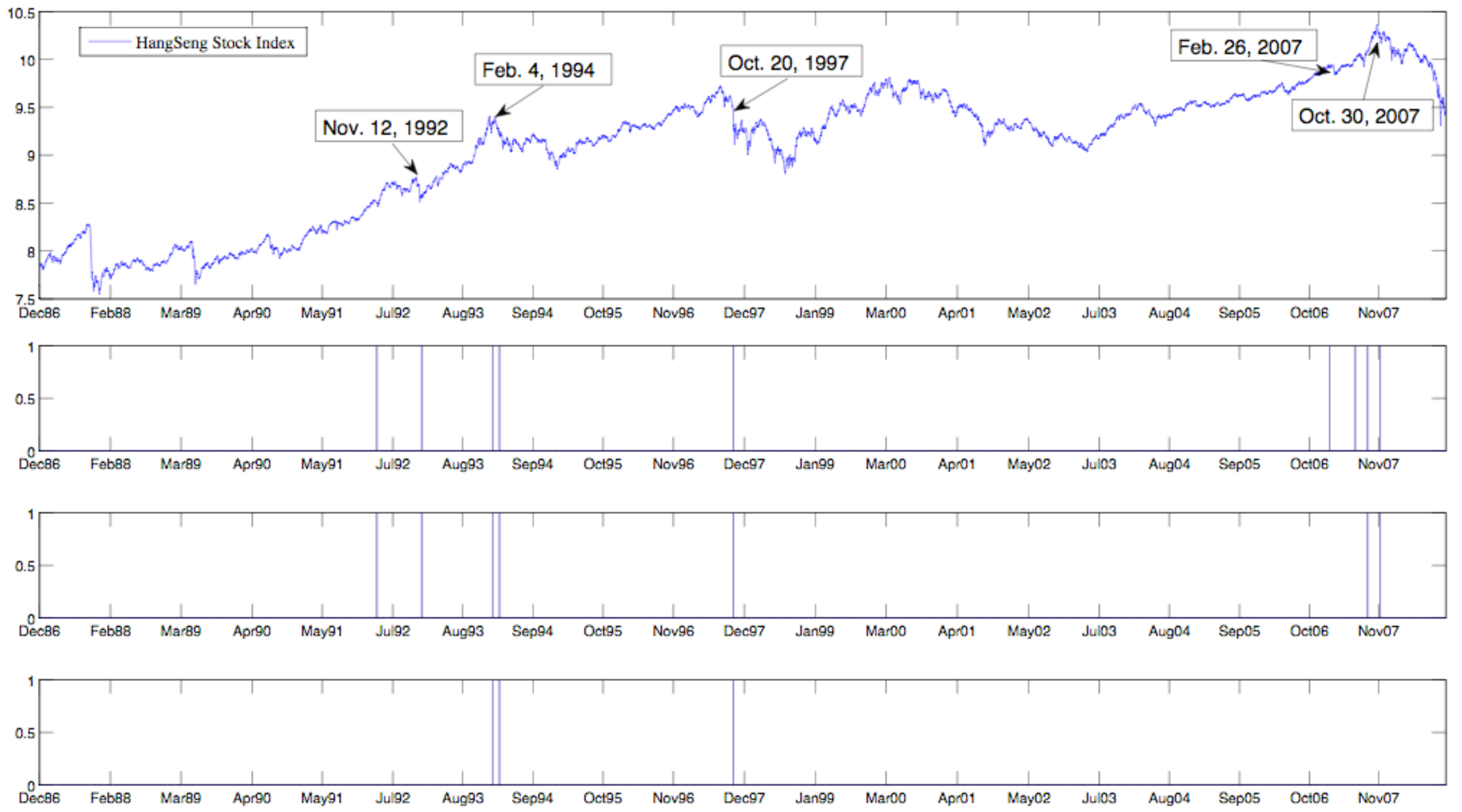}
\caption{Same as Fig.\protect\ref{F: 4} for the Heng Seng index of Hong Kong.} \label{F: 6}
\end{figure}

The results shown in Figs.\ref{F: 4}-\ref{F: 6} complement and refine those obtained 
with the bubble model tested in the previous subsection. In general, there are less alarms
when using this second model and procedure, compared with the first bubble model
and procedure of subsection \ref{S: 4.1}. One can observe in Fig.\ref{F: 4}
for the S\&P500 index two very well-defined clusters diagnosing a bubble ending 
with the crash of October 1987 and another bubble ending in October 2007.
The dates indicated in the upper panel, Oct. 17, 1987 and Oct. 9, 2007, correspond
to the right time of the last window in which an alarm is found for each of these two clusters.
The timing is thus remarkably accurate in terms of the determination of the
end of each bubble regime. It is interesting that this second model
in terms of a momentum bubble singularity is able to diagnose unambiguously a bubble
ending in October 2007, while the first model diagnosed only an intermediate phase
of this price development. This bubble can be referred to as the ``real-estate-MBS'' bubble
(MBS stands for morgage-backed security, \citet{SorWood09}). Using the level 1 filter
for the positive feedback exponent $m$, we observe in addition three false alarms.
Raising the condition that $m$ should be larger than $2.5$ (respectively $3$)
removes two (respectively all) of these false alarms.

Fig.\ref{F: 5} for the Nasdaq Composite index identifies the two bubbles ending in March 2000
(the ``new economy'' ICT bubble) and in Oct. 2007, without any false alarm.

Fig.\ref{F: 6} for the Heng Seng of Hong Kong similarly diagnoses these two bubbles 
as well as those ending with the 1992 and 1994 Asian events.

\section{Concluding remarks }

We have developed two rational expectation models of financial bubbles
with heterogeneous rational arbitrageurs. Two key ingredients characterize
these models: (i) the existence of a positive feedback quantified by a nonlinear
power law dependence of price growth as a function of either price or momentum;
(ii) the stochastic mean-reversion dynamics of the termination time of the bubble.
The first model characterizes a bubble as a faster-than-exponential accelerating stochastic price
ending in a finite-time singularity at a stochastic critical time.
The second model views a bubble as a regime characterized by an accelerating
momentum ending at a finite time singularity, also with at a stochastic critical time.
This second model has the additional feature of taking into account the existence
of some inertia in the price formation process, which is related to the momentum effect.

In these two models, the heterogeneous arbitrageurs exhibit distinct perception for 
the rising risk of a crash as the bubble develops. Each arbitrageur is assumed
to know the price formation process and to determine her exit time so as
to maximize her expected gain. The resulting distribution of exit times 
lead to a synchronization problem, preventing arbitraging of the bubble
and allowing it to continue its course up to close to its potential critical time.

The explicit analytical solutions of the two models allow us to propose
nonlinear transformations of the price time series into stochastic critical
time series. The qualification of a bubble regime then boils down
to characterize the nature of the transformed stochastic critical time series, thereby
avoiding  the difficult problem of parameter estimation of the stochastic differential 
equation describing the price dynamics. We develop an operational procedure
that qualifies the existence of a running bubble (i) if the critical time series
is found to reject a standard unit-root test at a high confidence level,
(ii)  if the exponent $m$ of the nonlinear power law characterizing the 
positive feedback is sufficient large and (iii) if the expected critical time
is not too distant from the time of the analysis.

The two procedures derived from the two bubble models
have been applied to  three financial markets, the US S\&P500 index from 1 February 1980 to 31 October 2008, 
the US NASDAQ composite index from 1 January 1980 to 31 July 2008 and the
Hong Kong Hang Seng index from 1 December 1986 to 30 November 2008.
Specifically, we have developed criteria to flag an alarm for the presence of a bubble,
that we validate by determining if the diagnosed bubble is followed by a crash in short order.
Remarkably, we find that the major known crashes over these periods are correctly
identified with few false alarms. The method using the second bubble model
in terms of a finite-time singularity of the price momentum seems to be more 
reliable with fewer false alarms and a better detection of the two principal bubbles
phases characterizing the last 30 years or so. 

These results suggest the feasibility of advance bubble warning.

\bigskip
\bigskip
\noindent{\bf Acknowledgments}
\noindent The authors would like to thank Stefan Riemann for useful discussions. 
We acknowledge financial support  from the ETH Competence Center ``Coping with Crises in Complex 
Socio-Economic Systems'' (CCSS) through ETH Research  Grant CH1-01-08-2.
 This work was partly supported by the National Natural Science Foundation of China for Creative Research Group: {\em Modeling and Management for Several Complex Economic System Based on Behavior } (Grant No.\,70521001). Lin Li also appreciates the China Scholarship Council (CSC) for supporting his studies at ETH Zurich (No.\, 2008602049).

\clearpage

\appendix
\section{Proof of Proposition 1}
\begin{proof}
Let $n=m-1$ and $z=p^{-n}$. Employing It\"{o} lemma for equation \eqref{E: stc1}, we have
\begin{align}
\df z&=\frac{\partial z}{\partial p}\df p+\frac{1}{2}\frac{\partial^2 p}{\partial p^2}(\df p)^2\\
&=-np^{-m}(\mu p^{m}[1+\delta(p,t)]dt+\sigma p^{m}dW)-\frac{1}{2}n(-n-1)p^{-m-1}\sigma^2 p^{2m}dt\\
&=-n\mu(1+\delta(p,t)-\frac{1}{2}(n+1)\sigma^2\mu p^{m-1})dt+n\sigma dW
\end{align}
Recalling that $\delta(p,t)=\alpha \tc+\frac{1}{2}m\mu \sigma^2 p^{m-1}$,
the previous expression can be simplified into 
\begin{align}
\df z &=-n\mu \df t+n\mu(-\alpha \tc \df t +{\sigma \over \mu} \df W)\\
&=-n\mu \df  t+n\mu\, \df \tc
\end{align}
Integrating both sides of the above equation and with $x(t=0)=p_0^{-n}$, we obtain
\begin{equation}
p=(n\mu)^{-\frac{1}{n}}[T_c+\tc-t]^{-\frac{1}{n}}, \qquad T_c=\frac{p_0^{-n}}{n\mu}~.
\end{equation}
Replacing $n$  by $\frac{1}{\beta}$, this reproduces the solution \eqref{6}.
\end{proof}

\section{Proof of Proposition 3}
\begin{proof}
We now check that equation \eqref{E: stcII} is the solution of the SDEs \eqref{stcII part1} and \eqref{stcII part2}. 
For this, we apply It\"{o} lemma on equation  \eqref{E: stcII} by regarding $\ln p_t$ as a function of $\tc$. This leads to
\begin{align}
\df \ln p&=\frac{\partial p}{\partial t}\df t+\frac{\partial p}{\partial \tc}\df \tc+\frac{1}{2}\frac{\partial^2 p}{\partial \tc^2}(\df\tc)^2\\
&=(1-\beta)B(T_c+\tc-t)^{-\beta}\df t-(1-\beta)B(T_c+\tc-t)^{-\beta}\df \tc\\
&\qquad+\frac{1}{2}\beta(1-\beta)B(T_c+\tc-t)^{-\beta-1}(\sigma / \mu)^2\df t
\end{align}
Taking into account that $B=\frac{1}{1-\beta}(\beta / \mu)^{\beta}$ and let $Z$ represent $(\beta / \mu)^{\beta}[T_c+\tc-t]^{-\beta}$, the above expression can be rewritten as
\begin{equation}
\df \ln p=Z \left[1+\alpha\tc+\frac{1}{2} (\sigma^2 / \mu) Z^{\frac{1}{\beta}} \right] \df t +Z (\sigma / \mu)\df W~.
\end{equation}
On the other hand, it is easy to see that $Z$ is the solution of \eqref{stcII part2} in the light of Proposition 1,
which leads to $Z=x$. Furthermore, we note that $\alpha\tc+\frac{\sigma^2}{2\mu} Z^{\frac{1}{\beta}}$ is
nothing but $\gamma(Z,t)$. Therefore $\ln p_t$ given by \eqref{E: stcII} satisfies  both \eqref{stcII part1} and \eqref{stcII part2}.
\end{proof}

\clearpage
\bibliographystyle{elsarticle-harv}
\bibliography{STCmodel}
\pagebreak

\clearpage

\clearpage

\clearpage

\clearpage

\clearpage

\end{document}